\pdfoutput=1

\documentclass[twocolumn,aps,prl,showpacs,superscriptaddress]{revtex4}

\usepackage{graphicx}
\usepackage{amsmath,amssymb}
\usepackage{color}
\usepackage[colorlinks=true,citecolor=blue,urlcolor=blue]{hyperref}

\begin{document}


\title{Simple Hardy-Like Proof of Quantum Contextuality}



\author{Ad\'an Cabello}
\affiliation{Departamento de F\'{\i}sica Aplicada II, Universidad de Sevilla, E-41012 Sevilla, Spain}

\author{Piotr Badzi{\c a}g}
\affiliation{Department of Physics, Stockholm University, S-10691, Stockholm, Sweden}

\author{Marcelo Terra Cunha}
\affiliation{Departamento de Matem\'atica, Universidade Federal de Minas Gerais,
 Caixa Postal 702, Belo Horizonte, Minas Gerais 30123-970, Brazil}

\author{Mohamed Bourennane}
\affiliation{Department of Physics, Stockholm University, S-10691, Stockholm, Sweden}


\date{\today}



\begin{abstract}
Contextuality and nonlocality are two fundamental properties of nature. Hardy's proof is considered the simplest proof of nonlocality and can also be seen as a particular violation of the simplest Bell inequality. A fundamental question is: Which is the simplest proof of contextuality? We show that there is a Hardy-like proof of contextuality that can also be seen as a particular violation of the simplest noncontextuality inequality. Interestingly, this new proof connects this inequality with the proof of the Kochen-Specker theorem, providing the missing link between these two fundamental results, and can be extended to an arbitrary odd number $n$ of settings, an extension that can be seen as a particular violation of the $n$-cycle inequality.
\end{abstract}



\maketitle


{\em Introduction.---}Contextuality (i.e., the impossibility of descriptions in terms of noncontextual hidden variables) and nonlocality (i.e., the impossibility of descriptions in terms of local hidden variables) are two fundamental properties of nature. Hardy's proof of nonlocality \cite{Hardy92,Hardy93} can be presented in very simple ways \cite{Mermin94a,Mermin94b,Mermin95,KH05}. Because of its simplicity, it is considered ``the best version of Bell's theorem'' \cite{Mermin95}.

The argument can be formulated in terms of four boxes which can be either empty or full. With $P(0,1|i,j)$ denoting the probability that box $i$ is empty and box $j$ is full (and likewise for both boxes full and both empty), one can write Hardy's conditions as
\begin{subequations}
\label{zero}
\begin{align}
 &P(1,1|1,4)=0, \label{zeroa}\\
 &P(1,1|2,3)=0, \label{zerob}\\
 &P(0,0|2,4)=0. \label{zeroc}
\end{align}
\end{subequations}
From these conditions, anyone who assumes that the result of finding the boxes empty or full is predetermined and independent of which other boxes are opened (i.e., noncontextuality of results) would conclude that $P(1,1|1,3)=0$. Nevertheless, in a physical experiment this implication can be violated. For that, one needs to prepare a suitable two-particle state and allow one observer to perform dichotomic measurements 1 and 2 on one of the particles and another observer to perform dichotomic measurements 3 and 4 on the other particle. With a suitable choice of measurements one can satisfy conditions (\ref{zero}) but violate the implication $P(1,1|1,3)=0$. For details, see \cite{Hardy92,Hardy93}.

In quantum description, an experiment testing Hardy's argument has to probe a four-dimensional Hilbert space and can be regarded \cite{Mermin94a} as an example for the violation of the simplest Bell inequality, the Clauser-Horne-Shimony-Holt inequality \cite{CHSH69}.

On the other hand, the simplest noncontextuality inequality violated by nature is due to Klyachko, Can, Binicio\u{g}lu, and Shumovsky (KCBS) \cite{KCBS08} and is violated already in three-dimensional quantum systems. A natural question is whether there is a Hardy-like proof of contextuality that may be seen as a violation of the KCBS inequality.


{\em Simple proof of quantum contextuality.---}Consider a physical system of five boxes, numbered from 1 to 5, such that each of them can be either empty or full. Let's denote as $P(0,1|2,3)$ the joint probability of finding box 2 empty and box 3 full.

One can prepare this system in a state such that
\begin{subequations}
 \label{uno}
\begin{align}
 &P(0,1|1,2)+P(0,1|2,3)=1, \label{unoa}\\
 &P(0,1|3,4)+P(0,1|4,5)=1. \label{unob}
\end{align}
\end{subequations}
Condition (\ref{unoa}) means that when box~2 is full then box~1 is empty and when box~2 is empty then box~3 is full. The condition can, thus, be reformulated as $P(1,1|1,2)=P(0,0|2,3)=0$. Similarly, condition (\ref{unob}) is equivalent to $P(1,1|3,4)=P(0,0|4,5)=0$.

From these conditions, anyone who assumes that the result of finding the boxes empty or full is predetermined and independent of which boxes are opened (i.e., noncontextuality of results) would conclude that
\begin{equation}
 P(0,1|5,1)=0.
 \label{dos}
\end{equation}

However, one can prepare a quantum system such that conditions (\ref{uno}) occur while (\ref{dos}) does not. For example, one can prepare a three-level quantum system (a qutrit) in the state
\begin{equation}
\label{state}
 |\eta\rangle = \frac{1}{\sqrt{3}}(1,1,1)^T,
\end{equation}
where $T$ means transposition and where opening box $i=1,\ldots,5$ is equivalent to measuring the projector on the states $|v_i\rangle$ given by
\begin{subequations}
\label{vec}
\begin{align}
 &|v_1\rangle = \frac{1}{\sqrt{3}}(1,-1,1)^T,\\
 &|v_2\rangle = \frac{1}{\sqrt{2}}(1,1,0)^T,\\
 &|v_3\rangle = (0,0,1)^T,\\
 &|v_4\rangle = (1,0,0)^T,\\
 &|v_5\rangle = \frac{1}{\sqrt{2}}(0,1,1)^T,
\end{align}
\end{subequations}
and empty and full are equivalent to obtain result 0 and 1, respectively.

Notice that the projectors onto $|v_1\rangle$ and $|v_2\rangle$ are compatible since $|v_1\rangle$ and $|v_2\rangle$ are orthogonal. Therefore, the joint probability of finding the result 0 for the projector onto $|v_1\rangle$ and the result 1 for the projector onto $|v_2\rangle$ for the state $|\eta\rangle$, denoted as $P_{|\eta\rangle}(0,1|1,2)$, is well defined, and the same happens for the other four probabilities in (\ref{uno}) and (\ref{dos}). Specifically,
\begin{subequations}
\begin{align}
 &P_{|\eta\rangle}(0,1|1,2)+P_{|\eta\rangle}(0,1|2,3)=\frac{2}{3}+\frac{1}{3},\\
 &P_{|\eta\rangle}(0,1|3,4)+P_{|\eta\rangle}(0,1|4,5)=\frac{1}{3}+\frac{2}{3}.
 \label{unoq}
\end{align}
\end{subequations}
However,
\begin{equation}
 P_{|\eta\rangle}(0,1|5,1)=\frac{1}{9},
 \label{dosq}
\end{equation}
in contradiction to (\ref{dos}).

Moreover, it can be easily shown that $\frac{1}{9}$ is the maximum value allowed by quantum theory (QT) for any system satisfying (\ref{uno}) \cite{Cabello94}.


{\em Connection with the KCBS inequality.---}The KCBS inequality and its maximum quantum bound can be written \cite{BBCGL11} as
\begin{equation}
 \sum_{i=1}^{5} P(0,1|i,i+1) \stackrel{\mbox{\tiny{ NCHV}}}{\leq} 2 \stackrel{\mbox{\tiny{Q}}}{\leq} \sqrt{5},
 \label{kcbs}
\end{equation}
where $5+1=1$, $\stackrel{\mbox{\tiny{ NCHV}}}{\leq} 2$ indicates that 2 is the maximum value for noncontextual hidden variable (NCHV) theories, and $\stackrel{\mbox{\tiny{Q}}}{\leq} \sqrt{5}$ indicates that $\sqrt{5} \approx 2.236$ is the maximum value in QT.

Clearly, the left hand side of (\ref{kcbs}) is nothing but the sum of the five probabilities in (\ref{uno}) and (\ref{dos}). For NCHV theories, this sum is upper bounded by~$2$. However, in our example, QT gives $2+\frac{1}{9}$. Therefore, the proof can be considered as a particular violation of the KCBS inequality.


{\em Connection with the Kochen-Specker theorem.---}Kochen and Specker were the first to prove the inconsistency between QT and NCHV theories \cite{Specker60,KS67} and did it using a construction involving 117 three-dimensional unit vectors which is traditionally illustrated using a graph in which vertices represent the vectors and adjacent vertices represent orthogonal vectors \cite{KS67}. This construction is obtained by replicating 15 times a basic building block which contains a set of eight vectors.

\begin{figure}[t!]
\vspace{-7.2cm}
\centering
\centerline{\includegraphics[scale=0.60]{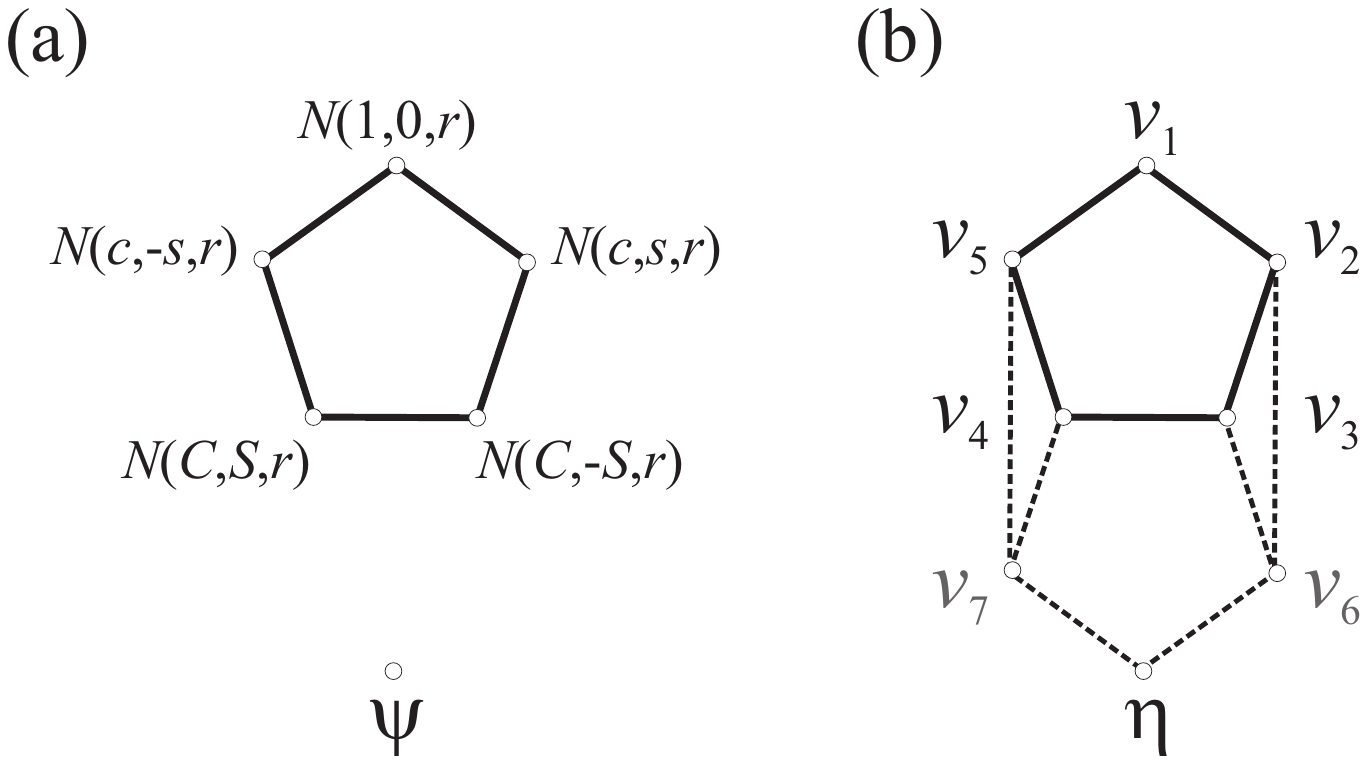}}
\vspace{-6.1cm}
\caption{\label{Fig1}(a) Graph of orthogonality between the vectors leading to a maximum quantum violation of the KCBS inequality for the state $|\psi\rangle=(0,0,1)^T$. $r=\sqrt{\cos{\left(\frac{\pi}{5}\right)}}$,
$c=\cos{\left(\frac{4\pi}{5}\right)}$, $s=\sin{\left(\frac{4\pi}{5}\right)}$,
$C=\cos{\left(\frac{2\pi}{5}\right)}$, $S=\sin{\left(\frac{2\pi}{5}\right)}$, and $N=1/\sqrt{1+r^2}$.
(b) Graph of orthogonality between the vectors in (\ref{state}) and (\ref{vec}) in the Hardy-like proof of quantum contextuality. The vectors that are not explicit in the Hardy-like proof but are explicit in the Kochen-Specker proof are $|v_6\rangle = \frac{1}{\sqrt{2}}(1,-1,0)^T$ and $|v_7\rangle = \frac{1}{\sqrt{2}}(0,1,-1)^T$.}
\end{figure}

On the other hand, the exclusivity (orthogonality) relationships between the five events (vectors) in (\ref{kcbs}) are represented by a pentagon. For example, a choice of vectors leading to the maximum quantum violation of the KCBS inequality is represented in Fig.~\ref{Fig1} (a). Notice that the same relationships occur for the Hardy-like proof of quantum contextuality. Moreover, because of the additional requirement (\ref{unoa}), there must exist a vector $|v_6\rangle$ that is orthonormal to $|\eta\rangle$, $|v_2\rangle$, and $|v_3\rangle$, and because of requirement (\ref{unob}), there must exist a vector $|v_7\rangle$ that is orthonormal to $|\eta\rangle$, $|v_4\rangle$, and $|v_5\rangle$. If we represent the orthogonality relationships of these eight vectors [see Fig.~\ref{Fig1} (b)], we end up with exactly the basic eight-vector set of the Kochen and Specker proof.


{\em Generalization to an arbitrary number of settings and connection with the odd cycle inequalities.---}Hardy's proof can be extended to multiple settings \cite{Hardy97,BBDH97,BDDM05} and the resulting proof is a particular violation of the chained Bell inequalities \cite{BC89a,BC89b,BC90}, which have many fundamental applications \cite{AKLZ99,Peres00,BHK05,BKP06,CR08,CR11,CR12}. A natural question is whether a similar extension is possible for the contextuality proof.

Consider a physical system of an odd number $n \ge 7$ of boxes, numbered from 1 to $n$, such that each of them can be either empty or full. One can prepare this system in a state such that
\begin{subequations}
 \label{nuno}
\begin{align}
 &P(0,1|1,2)+P(0,1|2,3)=1, \label{nunoa}\\
 &P(0,1|3,4)+P(0,1|4,5)=1,\ldots, \label{nunob}\\
 &P(0,1|n-2,n-1)+P(0,1|n-1,n)=1.\label{nunob}
\end{align}
\end{subequations}
From these conditions, anyone assuming noncontextuality of results would conclude that
\begin{equation}
 P(0,1|n,1)=0.
 \label{ndos}
\end{equation}

\begin{figure}[t!]
\vspace{-6.7cm}
\centering
\centerline{\includegraphics[scale=0.58]{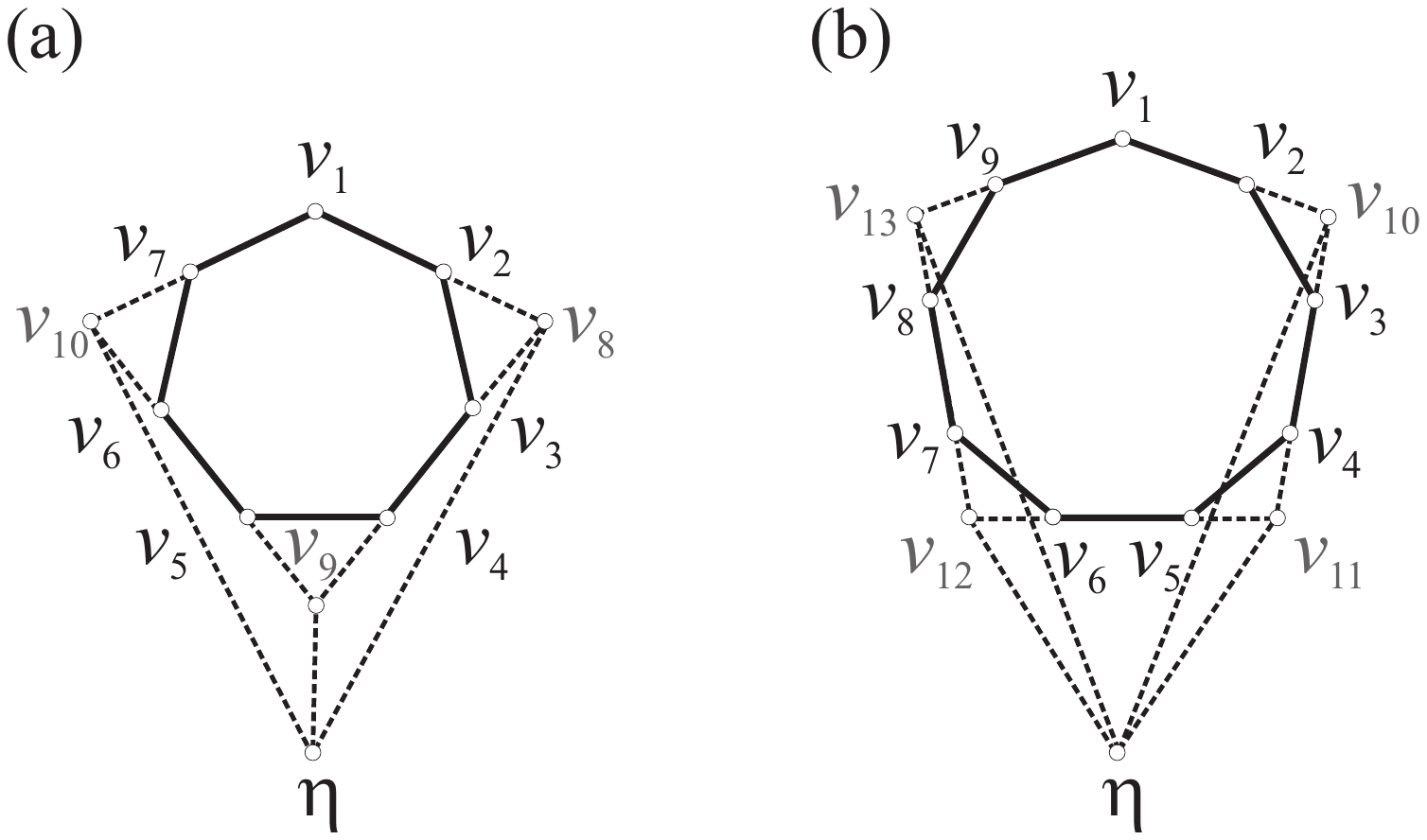}}
\vspace{-5.4cm}
\caption{\label{Fig2}(a) Graph of orthogonality between the vectors in the Hardy-like proof of quantum contextuality with 7 settings. (b) Idem with 9 settings.}
\end{figure}

However, for any odd $n \ge 7$, one can prepare a qutrit such that conditions (\ref{nuno}) occur while (\ref{ndos}) does not. Curiously, the orthogonality graph for the case $n=7$ shown in Fig.~\ref{Fig2} (a) was first studied in Ref.~\cite{Bell66}. The maximum quantum value for $P(0,1|7,1)$ is $\frac{1}{5}$ \cite{CG96}. The case $n=9$ shown in Fig.~\ref{Fig2} (b) was first studied in Ref.~\cite{CG95} and the maximum quantum value for $P(0,1|9,1)$ is $\left(1+[16/3 \sqrt{3}]\right)^{-1}$ \cite{CG95}. The maximum quantum value for $P(0,1|n,1)$ tends to $1/2$ when $n$ tends to infinity (the same value as in Hardy's ``ladder'' proof \cite{Hardy97,BBDH97,BDDM05}); see the Appendix for a proof.

Interestingly, this extended contextuality proof is a particular violation of a generalization of the KCBS inequality to an odd number $n \ge 7$ of settings. These generalized inequalities were introduced, independently, in Refs.~\cite{LSW11,CSW10}, and are called odd cycle inequalities. For any $n \ge 5$ odd,
\begin{equation}
 \sum_{i=1}^{n} P(0,1|i,i+1) \stackrel{\mbox{\tiny{ NCHV}}}{\leq} \frac{n-1}{2} \stackrel{\mbox{\tiny{Q}}}{\leq} \frac{n \cos\left(\frac{\pi}{n}\right)}{1+\cos\left(\frac{\pi}{n}\right)},
 \label{nkcbs}
\end{equation}
where $n+1=1$. The case $n=5$ corresponds to the KCBS inequality (\ref{kcbs}). A number of arguments regarding why these inequalities are fundamental tests of QT and the experimental settings leading to the maximum quantum violation of the inequalities can be found in Ref.~\cite{CDLP12}. The noncontextuality polytopes associated with these inequalities are fully characterized in Ref.~\cite{AQBTC12}.


{\em Conclusion.---}In this Letter we generalize the reasoning behind Hardy's proof onto single systems realizable in a three-dimensional Hilbert space. Our argument provides a particularly simple realization of quantum contextuality and links two fundamental results: the original (state-independent and inequality-free) proof of impossibility of NCHV theories in QT \cite{KS67} and the simplest experimentally testable noncontextuality inequality \cite{KCBS08,AACB13}. Moreover, the proof can be extended to an arbitrary odd number $n$ of settings, and this extension provides a particular example of the violation of the $n$-cycle inequalities for any odd $n$ and connects these recently discovered inequalities with previous proofs of quantum contextuality.


{\em Appendix.---}Equations.~(\ref{nuno}) and (\ref{nkcbs}) imply that $P(0,1|n,1) \leq \frac{1}{2}$. Here we show that $\lim _{n\rightarrow \infty} P(0,1|n,1) = \frac{1}{2}$. For that, it is enough to give a state $|\eta\rangle$ and a set of projectors such that $\lim _{n\rightarrow \infty} P_{|\eta\rangle}(0,1|n,1) = \frac{1}{2}$. For simplicity, we restrict the discussion to the case $n = 4k+1$.

Let measurement $i$ be the projector on state $|i\rangle$ and let $|1\rangle, |2\rangle,\ldots,|4k\rangle, |4k+1\rangle$ be given by
\begin{subequations}
\label{eq:statesforp12}
\begin{align}
 |1\rangle \propto & (0, \cos \theta_1, \sin \theta_1 \cos\phi_1)^T,\\
 |2j\rangle =& (-\cos \theta_j \sin \phi_j, \cos \theta_j \cos \phi_j, \sin \theta_j)^T,\\
 |2j+1\rangle =& (\sin \theta_j \sin \phi_j, -\sin \theta_j \cos \phi_j, \cos \theta_j)^T,\\
 |4k-2j+2\rangle =& (-\sin \theta_j \sin \phi_j, -\sin \theta_j \cos \phi_j, \cos \theta_j)^T,\\
 |4k-2j+3\rangle =& (\cos \theta_j \sin \phi_j, \cos \theta_j \cos \phi_j, \sin \theta_j)^T,
\end{align}
\end{subequations}
for $j = 1, \ldots, k$.
The choice of these vectors is motivated by the sharing of one common direction for all planes generated by neighbor pairs. Explicitly, the state
\begin{equation}
 \label{eq:etaforp12}
 |\eta\rangle = (0,0,1)^T
\end{equation}
can be written as
\begin{subequations}
\begin{equation}
 |\eta\rangle = \sin \theta_j |2j\rangle + \cos \theta_j |2j+1\rangle,
\end{equation}
and also as
\begin{equation}
 |\eta\rangle = \cos \theta_j |4k-2j+2\rangle + \sin \theta_j |4k-2j+3\rangle,
\end{equation}
implying Eqs.~(\ref{nuno}).
\end{subequations}

The vectors in Eq.~\eqref{eq:statesforp12} fulfill the orthogonality relations $\left\langle 2l \middle| 2l+1 \right\rangle = 0$, for $1 \leq l \leq 2k$.
For the other compatibility conditions to be satisfied, we must also demand that $\left\langle 2l-1 \middle| 2l \right\rangle=0$, for $1\leq l\leq 2k$ and $\left\langle 4k+1\middle| 1\right\rangle = 0$ which, except by $l= 2(k+1)$, all give
\begin{subequations}
\label{eqrecp12}
\begin{equation}
 \tan \theta_{j+1} \cos \Delta_j = \tan \theta_j,
\end{equation}
where $\Delta_j = \phi_{j+1} - \phi_j$ and $1\leq j < k$;
the exceptional case $l= 2(k+1)$ gives
\begin{equation}
 \tan^2 \theta_k = -\cos 2\phi_k.
\end{equation}
\end{subequations}

If one chooses $\phi_j = \frac{j\pi}{2(k+1)}$, all $\Delta_j = \frac{\pi}{2(k+1)}$ and Eqs.~\eqref{eqrecp12} define all $\theta_j$ such that all the vectors are distinct.

Under the above choices,
\begin{equation}
 P(0,1|n,1) = \frac{\sin^2 \theta_1 \cos^2\phi_1}{{\cos^2 \theta_1+\sin^2 \theta_1 \cos^2\phi_1}},
\end{equation}
and, as $k \rightarrow \infty$, one has
\begin{equation}
 P(0,1|n,1) \sim \sin^2\theta_1 \rightarrow \frac{1}{2}.
\end{equation}


\begin{acknowledgments}
 We thank Jaewan Kim and Gl\'aucia Murta for stimulating discussions. This work was supported by Project No.\ FIS2011-29400 (MINECO, Spain), the Brazilian agencies Fapemig, Capes, and CNPq, the Swedish Research Council (VR), and the ERC Advanced Grant QOLAPS. A.\,C. thanks M.\,B. for his hospitality at Stockholm University.
\end{acknowledgments}


\end{document}